\documentstyle[prb,aps,multicol]{revtex}

\begin{document}
\title{Tunneling  in a cavity\footnote{appears in The Physical Review A}}

\author{Peter Neu  and  Robert J. Silbey}
\address{Department of Chemistry, Massachusetts Institute of Technology,
 Cambridge, Ma.  02139}

\date{\today}
\maketitle
\thispagestyle{empty}

\begin{abstract}
{\bf Abstract.}
The mechanism of coherent destruction of tunneling  found by
Grossmann {\it et al.} [Phys. Rev. Lett. {\bf 67}, 516 (1991)]
is studied from the viewpoint of quantum optics by considering 
 the photon statistics of a single mode cavity field
which is strongly coupled to a  two-level tunneling system (TS).
As a function of the interaction time between TS and cavity 
the photon statistics displays the tunneling dynamics.
 In the semi-classical limit of high photon
occupation number $n$,  coherent destruction of tunneling is exhibited 
in  a slowing down of an  amplitude modulation for  certain parameter
ratios of the field.
  The phenomenon  is  explained as arising from
interference between displaced number states in phase space which survives
the large $n$ limit due to  identical  $n^{-1/2}$ scaling between orbit width 
and displacement.

\vspace{0.2cm}
\noindent
PACS number(s): 42.50.Ct, 42.50.Lc, 05.30.-d
\end{abstract}




\begin{multicols}{2}

In recent years the  idea of modulation and therefore control of  tunneling
 by a monochromatic electromagnetic field 
has been subject of considerable interest. 
The typical Hamiltonian describes a particle in an isolated 
 double-well potential (DWP) which is 
periodically driven by an external  force
\begin{equation} \label{cl}
H(t) = H_{\rm DWP}(x) + Sx\cos(\omega_L t).
\end{equation}
Here $H_{\rm DWP}(x)$ is the Hamiltonian of the DWP,
  $S$ is the amplitude and  $\omega_L$ the driving frequency.
The attention has mainly
focused on a possible enhancement or suppression of coherent tunneling:
Lin and Ballentine\cite{LB} have demonstrated that the tunneling probability
is highly enhanced due to periodic modulation for high field strengths and 
driving frequencies close to the classical 
oscillation frequency at the bottom of each well.
In the opposite limit Grossmann, H\"anggi and coworkers\cite{5} found complete
suppression of tunneling such that a particle initially localized in one of 
the two wells will never escape to the other well. They termed this effect
``coherent destruction of tunneling''. 
It has since been of continuing  interest.\cite{TLS1,TLS2,12,11,E1,E2,10}
The most surprising feature  
is the {\it periodicity} of  the destruction of tunneling for certain parameter
ratios of  $S$ and  $\omega_L$.
So far there is no clear understanding of this phenomenon.

By using the Floquet formalism it has been shown 
that many characteristic features of the tunneling suppression can already
be described in a two-level approximation of the DWP.\cite{TLS1,TLS2}
 Many different aspects 
of the effect have been illuminated in this framework: Makarov\cite{12},
Plata and  Gomez Llorente\cite{11} quantized the electromagnetic field
and recovered the effect in the limit of large number of photons in the field.
Wang and Shao\cite{E1} mapped the driven two-level dynamics to a classical
one of a charged  particle moving in a harmonic potential plus a magnetic field 
in a plane. Kayanuma\cite{E2} explained the suppression of tunneling as an effect arising 
from interference at periodic level crossings.

In the present Rapid Communication,
an alternative explanation is proposed which uses the concept 
of {\it phase space interference}  known from squeezed states\cite{4b,8}
and displaced number states\cite{7a,7b} in quantum optics.

Let us start with a description of the physical situation we have in mind. At $t=0$ a
 single  two-level tunneling system (TS) prepared in a state localized in say the 
left well  $|L\rangle$
is injected in a cavity and starts to tunnel between its left and right state $|L\rangle$
and $|R\rangle$.
 The cavity contains a single mode which has been prepared in a 
number state $|n\rangle$.  We
 consider an ideal cavity, i.e., we neglect any kind
of dissipation. After an interaction time $t_{in}$ the TS leaves the cavity 
and the photon number distribution is measured irrespective of the state of the TS.

Thus our aim will be to calculate the transition probability from  the product state
$|L,n\rangle = |L\rangle |n\rangle$ at $t=0$ to another product state $|i,l\rangle$
for any $i = L,R$ at time $t = t_{in}$,
\begin{equation} \label{1}
P_{ln}(t_{in}) = \sum_{i=L,R }|\langle i,l|\Psi(t_{in})\rangle |^2
\end{equation}
where $|\Psi(t_{in})\rangle = e^{-iHt_{in}/\hbar}|L,n\rangle$.
For fixed $t_{in}$  this is the photon number distribution with $P_{ln}(0) = \delta_{ln}$.
For fixed $l$, $P_{ln}(t_{in})$ describes the dynamics of the cavity mode
interacting with a TS.

An important aspect of the present problem is the strong coupling
between the cavity mode and  the TS, and the separation of timescales 
between the {\it slow} tunneling motion and the {\it fast} field oscillations.
If we denote the coupling energy by $g$  and the tunneling frequency 
by $\Delta$, this means that we are interested
in the limit $\Delta\ll\omega_L$ and $g\sim\hbar\omega_L$ for low, and
$\sqrt{n} g\sim\hbar\omega_L$ for high number of photons in the field.

In this regime   the field strongly dresses 
the TS and both TS + field have to be treated as a single unit.
In contrast to the situation where $\omega_L \approx \Delta \gg g/\hbar$,
 we are confronted in the present case 
with a situation  in which the rotating-wave approximation  is not applicable. Thus instead
of using  the Jaynes-Cummings Hamiltonian\cite{6}, we must include counter-rotating terms
so that our Hamiltonian reads
\begin{equation} \label{2}
     H = -\frac{\hbar\Delta}{2} \sigma_x + g \sigma_z (a^{\dagger} + a)
       + \hbar\omega_L a^{\dagger}a - {g^2\over\hbar\omega_L}
\end{equation}
where $a^{\dagger}$, $a$ are the bosonic creation and annihilation operators
of the cavity mode. We have identified
$\sigma_x = |L\rangle\langle R| + |R\rangle\langle L|$ and $\sigma_z = 
|L\rangle\langle L| - |R\rangle\langle R|$.
Expressing the spin operator in the eigenstates of the TS,
$|0\rangle = 2^{-1/2}(|L\rangle + |R\rangle$ and 
$|1\rangle = 2^{-1/2}(|L\rangle - |R\rangle$,
the usual formulation
of the Hamiltionian in quantum optics is recovered.
We also have added a constant energy shift for later convenience. 
It has to be noted that suppression of tunneling does not occur
in the Jaynes-Cummings model.\cite{11} Hence the situation here
 is fundamentially different from the one usually encountered  in quantum optics.

Owing to the separation of timescales between the tunneling and the oscillation dynamics,
$P_{ln}(t_{in})$ can be calculated in perturbation theory in $\Delta/\omega_L$ by 
introducing dressed states
\begin{equation} \label{3}
|j(n)\rangle = U |n\rangle {1\over\sqrt{2}} ( |L\rangle + (-)^j |R\rangle)
\end{equation}
with $j = 0,1$ and 
\begin{equation} \label{4}
U = \exp[-\sigma_z \alpha (a^{\dagger} - a)]
\end{equation}
where $\alpha = g/\hbar\omega_L$. Equivalently, one may perform the polaron transformation
$H\rightarrow U^{\dagger}H U$ and continue to use the product state instead of the dressed
state basis. In the dressed state  basis the Hamiltonian can be written as 
\begin{eqnarray}\label{5}
  H &=& H_D + V \nonumber\\
     &=& \sum_{m,j}|j(m)\rangle ( m\hbar\omega_L - (-)^j {\hbar\Delta_m\over 2})\langle j(m)|
        \nonumber\\
    &-& \frac{\hbar\Delta}{4}\sum_{m,m',j,j'} |j(m)\rangle ( (-)^{j'} 
         (1 + (-)^{j-j'+m-m'}) 
       \nonumber\\
      &&\qquad\quad \times \  D_{mm'}(2\alpha)  (1 - \delta_{jj'}\delta_{mm'}))
         \langle j'(m')|
\end{eqnarray}
with
\begin{equation}\label{6}
\Delta_n = \Delta  D_{nn}(2\alpha)
\end{equation}
and 
\begin{eqnarray} \label{7}
    D_{ln}(\alpha) &=& \langle l|D(\alpha)|n\rangle \nonumber\\
                   &=& \left({n!\over l!}\right)^{1\over 2} \alpha^{l-n} e^{-{1\over 2} |\alpha|^2}
                       L_n^{l-n}(|\alpha|^2)
\end{eqnarray}
where $D(\alpha)= \exp[\alpha(a^{\dagger} - a)]$ is the shift operator and 
$L_n^{l-n}(x)$ an associate Laguerre polynomial ($l \ge n$).
The diagonal part $H_D$ 
builds a ladder of tunneling doublets with  intra-doublet spacing $\hbar\Delta_m$
and inter-doublet spacing $\hbar\omega_L$.  Because $\langle 1(m)|V|0(m)\rangle = 0$, 
$V$ induces only  mixing between dressed states  belonging to {\it different} doublets.
Hence, corrections to the dressed states are $O(\Delta/\omega_L)$. Neglecting 
$V$ for this reason we find for the transition probability  if $l \ge n$
\begin{eqnarray} \label{8}
    && P_{ln}(t_{in}) = \left|\sum_m D_{lm}(\alpha) D_{mn}(-\alpha)
                         e^{-im\omega_Lt_{in}}
                 \cos\left({\textstyle {1\over 2}}\Delta_m t_{in}\right)\right|^2
     \nonumber \\
                 &&\qquad   +\  \left|\sum_m D_{lm}(\alpha) D_{mn}(\alpha) 
                        e^{-im\omega_L t_{in}} 
            \sin\left({\textstyle {1\over 2}}\Delta_m t_{in}\right)\right|^2   
\end{eqnarray}
and $P_{ln}(t_{in}) = P_{nl}(t_{in})$ if $l < n$.
{\it We conclude that the tunneling dynamics can be seen in the spectrum of 
 the transition probability of the cavity field.}  
In addition to harmonics of $\omega_L$ its power spectrum also contains 
resonances at $\Delta_m$ arising from
the tunneling motion.
This behavior strongly depends
on the initial preparation of the TS. Injecting it in its ground state 
$|0\rangle $, for instance, yields only
resonances at $m\omega_L + (-)^\xi {1\over 2} \Delta_m$ where $\xi = 0,1$ depending
on whether $n-m$ is even or odd, respectively, and hence only rapid oscillations.

The effect of a strong coupling between the cavity mode and the TS is to mix a 
coherent amplitude  $\alpha \equiv g/\hbar\omega_L$ with 
 the intial number state of the mode.
This happens 
by displacing the oscillator wave function $\varphi_n(x) = \langle x|n\rangle$ like
$x \to x - \sqrt{2} \alpha$ in dimensionless  coordinates 
$x = (\mu\omega_L/\hbar)^{1/2}q$. 
Thus by injecting a TS which strongly couples to the cavity field, it  is  possible
to realize displaced number states\cite{7a,7b}.
  The statistical properties of displaced
number states have been discussed in Ref. \CITE{7b}. The photon number distribution 
is simply given by $P_{\rm DNS}(l) = |D_{ln}(\alpha)|^2$
 because $D_{ln}(\alpha)$ is the probability
amplititude of finding $l$ photons in a displaced number state $|\alpha, n\rangle =
D(\alpha) |n\rangle$. 

 The photon distribution (\ref{8}) 
for number states which are {\it dynamically} displaced
by the tunneling process resembles $P_{DNS}(l)$.
The displaced number states can either be shifted
into the same well (first term in (\ref{8})) or opposite wells (second term in (\ref{8})).
 As expected for $\alpha = 0$, $P_{ln}(t_{in}) = 
\delta_{ln}$ independent of the interaction time.

 In Fig. 1(a),(b) we have plotted 
$P_{ln}(t_{in})$ for $n=10$, $\omega_L t_{in} = 250$ and $\alpha =$ 0.25 and 0.5.
Whereas a coherent state with $\bar{n} = 10$ obeys the familiar Poisson distribution\cite{8}
shown by the line, the dynamically displaced number states exhibit  oscillations shown
by the histogram. Increasing $\alpha$ results in further oscillations.
This effect is independent of the specific value of $\omega_L t_{in}$ for $t_{in}\neq 0$,
though the absolute value of $P_{ln}(t_{in})$ depends on $\omega_L t_{in}$.
These modulations are clearly exhibited in the asymptotic expansion of (\ref{8}) in  the 
semi-classical limit (Bohr's correspondence principle):
 $l,n\to \infty$, $n/l\to 1$ with $l-n\ge 0$ finite. If we scale the coupling constant
between the field and the TS as $g\propto n^{-1/2}$ and note that associate Laguerre
polynomials asymptotically approach Bessel functions\cite{9} one finds in the semi-classical
limit 
\begin{eqnarray} \label{9}
      P_{l-n}(t_{in}) &=& \cos^2\left({\textstyle {1\over 2}}\widetilde{\Delta} t_{in}\right)\,
                          J_{l-n}^2\left({2\Omega\over\omega_L} \sin(\omega_Lt_{in})\right)
\nonumber \\
                    &+&   \sin^2\left({\textstyle {1\over 2}}\widetilde{\Delta} t_{in}\right)\,
                          J_{l-n}^2\left({2\Omega\over\omega_L} \cos(\omega_Lt_{in})\right).
\end{eqnarray}
Here 
\begin{equation}\label{10}
\widetilde{\Delta} = \Delta \, |J_0(2\Omega/\omega_L)|
\end{equation}
is the renormalized tunneling frequency, and 
\begin{equation}\label{11}
\Omega = 2\sqrt{n} g/\hbar
\end{equation}
is the Rabi frequency. In Fig. 1(c),(d), Eq. (\ref{9})
is plotted for $\Omega/\omega_L =$ 1 and 3.3. It shows the same  oscillations as the 
exact expression (\ref{8}). 

The expression (\ref{9}) is easily understood. The cosine and the sine factors 
represent the probability for 
 the tunneling particle to stay in the well where it has been prepared 
initially, or to escape to the other well, respectively.\cite{TLS1} The Bessel functions
represent the probability for the corresponding displaced number state to contain
$l$ photons if it  had $n$ initially (note that 
$P_{\rm DNS}(l)   \to J_{l-n}^2(\Omega/\omega_L)$
in the semi-classical limit). Hence the Bessel functions represent 
the photon statistics of displaced number states with an effective displacement
$(2\Omega/\omega_L)\sin(\omega_L t_{in})$ or $(2\Omega/\omega_L)\cos(\omega_L t_{in})$. 
From (\ref{10}) and the cosine and sine factors in (\ref{9}) we further  see that 
 for specific parameter values of the driving field -- where $2\Omega/\omega_L$
hits the roots of the zero order 
Bessel function  -- the tunneling process
is completely suppressed.
Note that localization cannot occur for small photon number in the field
because there is no value of $\alpha$ which is simultaneously a root of all Laguerre
polynomials. 
We conclude that in the large $n$ limit {\it the tunneling dynamics 
is displayed in the photon statistics by an amplitude modulation of the cavity mode oscillations.}
{\it  Coherent destruction of tunneling   manifests itself in
 the slowing down of this amplitude modulation} as
depicted in Fig. 2. 

We now show how the present picture can give a quantitative  understanding of this effect.
First recall that the oscillation in the photon number statistics in Fig. 1 originate
from   phase space   interference:\cite{7b}   the $n$th
number state can be associated with 
a circular band of width $(2n)^{-1/2}$ around its orbit with radius 
$\sqrt{2(n+1/2)}$ and centered around the origion.
Analogously so  can the $l$th displaced  number state which
is shifted by $\sqrt{2}\alpha$.  According to the area-of-overlap
concept\cite{4b}, the transition amplitude between two states is
 governed by the sum of all possible 
overlap areas weighted with appropriate phases. This results in the interference 
between contributions from different overlaps which can be constructive or destructive,
thus, giving rise to oscillations in the photon number distribution.
Another way of understanding the oscillations  is that  they
 arise from the possibility of a constructive and destructive
overlap between  displaced harmonic oscillator wave functions by
noting  that $D_{lm}(\alpha)$ is a polynomial in $l$ of order min$(l,m)$, i.e., has
min$(l,m)$ roots. 

From this consideration it becomes apparent that varying the relative 
radius of the two bands by changing $l$ vs. $n$, or varying the relative 
displacement by changing $\alpha$ for fixed $l$ and $n$ 
should result in similar effects.
If one notices further that the factor $J_0(2\Omega/\omega_L)$ in (\ref{10})
 is the probability amplitude
of finding in a number state   displaced by 
$x \rightarrow x - 2{\Omega/\omega_L \over \sqrt{2n}}$
 again exactly
the same number of photons, one expects that  the periodic suppression of tunneling
arises from  destructive and constructive 
interference  of displaced harmonic oscillator wave functions
in phase space.

To verify this argument, the $n\to \infty$ limit has to be considered with care.
 First take the number of photons in the cavity mode
to be finite. The factor $D_{nn}(2\alpha)$ which scales the doublet splitting (\ref{6})
is the overlap  between a  number state $|n\rangle$ shifted into the right well with the same
number state shifted into the left well.
If we note that $D_{nn}(2\alpha)$ is a polynomial in $|\alpha|^2$ of degree $n$,
and consequently has $n$ zeros, 
we expect $n+1$ oscillations of  $D_{nn}(2\alpha)$ as a function of 
$\alpha$. The $n$ zeros between the maxima result from  the $n$ possible ways
of destructive overlap between displaced  harmonic oscillator wave functions.
This explains Fig. 2 in Ref. \CITE{12}. However, contrary to the claim there,
this picture remains also valid if the oscillator energy is larger than the
reorganization energy $(n+{1\over 2})\hbar\omega_L > g^2/\hbar\omega_L$, and even in 
the limit $n\to \infty$. The simple reason for this is that in the semi-classical limit
$n\to \infty$ with  $2\sqrt{n}g/\hbar\to \Omega$
 fixed, both the distance between 
the nodes of the harmonic oscillator wave function (as a function of $x$) {\it and}
the displacement $x \rightarrow x - 2{\Omega/\omega_L \over \sqrt{2n}}$
scale as $n^{-1/2}$. Based on this argument, we expect oscillation to become noticable
as soon as the displacement exceeds the bandwidth, i.e.,  $2\Omega/\omega_L>1$.
Finally, we note that with the scaling chosen above both displaced orbits will always
intersect no matter how large $\Omega/\omega_L$ is. This  results in the infinite
number of oscillations seen in  the zero order Bessel function (\ref{10}).

Summarizing coherent destruction of tunneling  is explained 
 as a {\it quantum effect arising
from the destructive interference  of displaced harmonic oscillator wave functions
in phase space}.
The effect is strictly speaking only observable in the semi-classical limit 
for the reasons
mentioned above, and survives the large $n$ limit since both
 {\it the bandwidth of the orbit 
 and the displacement scale in the same way to zero as} $n\to\infty$.

In closing, we note that experimental study of the effects described here
is presently still out of range.
In a microcavity the driving frequency $\omega_L$ is $O$(GHz) whereas
typical values of the one-photon Rabi frequency $g/\hbar$ are $O$(kHz).\cite{1}
Hence a number state with large $n$ is needed which is difficult to realize
experimentally. Possibly a trap is more suited because of
its lower  driving frequency  $\omega_L\sim O$(MHz).\cite{Wal}
Damping  of the cavity mode which is too strong 
will also render an  observation impossible.
Furthermore the tunneling dynamics must still be coherent.
Finally the  superposition state $|L\rangle
= 2^{-1/2}(|0\rangle + |1\rangle)$ is difficult 
to prepare because the particle experiences  strong electromagnetic 
fields when it enters and leaves the cavity\cite{Wal} (for a discussion of this point
if one uses quantum wells, see Ref. \CITE{Met}).

$\\$
$\\$
This research has been supported in part by the NSF and by the 
Alexander von Humboldt Foundation. We also thank Prof. Peter H\"anggi
and Prof. Herbert Walther, as well as Dr. Min Cho
and Dr. Jochen Rau for discussions.

\section*{Figure captions}
\begin{itemize}
\item[FIG. 1.]    Characteristic  oscillations in the 
                  photon number distribution of dynamically 
                  displaced number states for $\omega_L t_{in} = 250$,
                  $\Delta/\omega_L = 0.2$ and 
                  $|n\rangle = |10\rangle$. In (a) and (b) the quantum limit, Eq. (\ref{8}),
                  is displayed for different values of  $\alpha \equiv g/\hbar\omega_L$. 
                  In (c) and (d) the semi-classical limit, Eq. (\ref{9}), is shown for 
                  different values of $\Omega/\omega_L \equiv 2\sqrt{n}\alpha$.
                  The full line shows the Poisson distribution with $\bar{n} = 10$.
                  
\item[FIG. 2.]    Coherent destruction of tunneling monitored in the
                  cavity field. Displayed is Eq. (\ref{9}) for $l=n$, 
                  $\Delta/\omega_L = 0.2$ as a function
                  of $\omega_L t_{in}$: upper figure   $2\Omega/\omega_L = 2$,
                  lower figure  $2\Omega/\omega_L = 2.3$. The first root of 
                  $J_0(2\Omega/\omega_L)$ occurs at $2\Omega/\omega_L \approx 2.405$.
                  The effect is exhibited by a decrease 
                  of the amplitude modulation in the cavity mode oscillations. 
\end{itemize}         

\end{multicols}

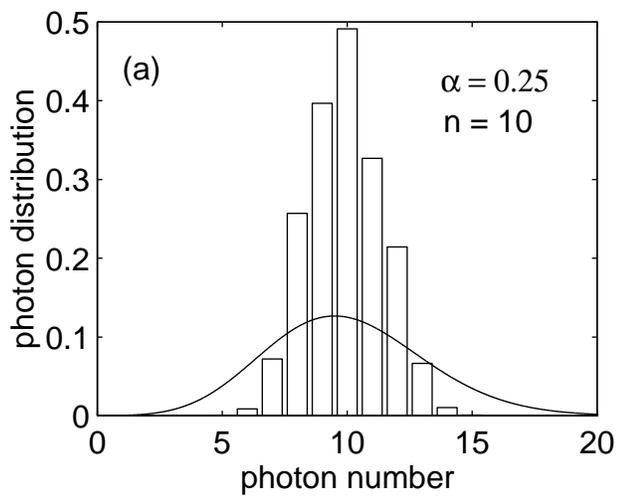
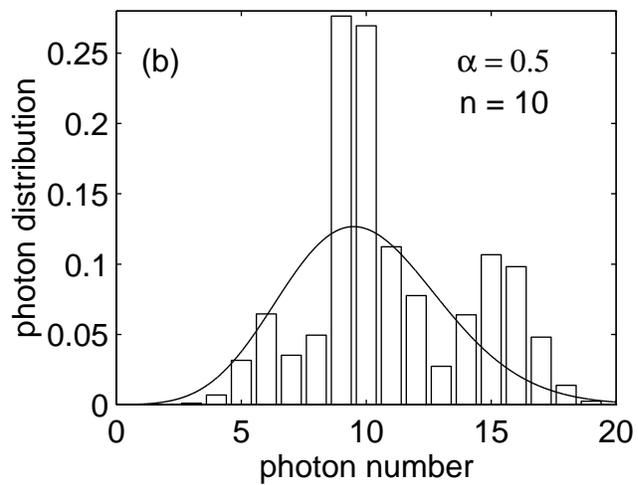
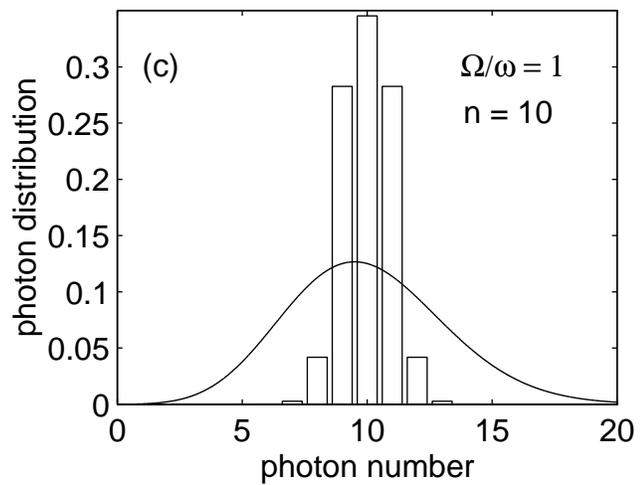
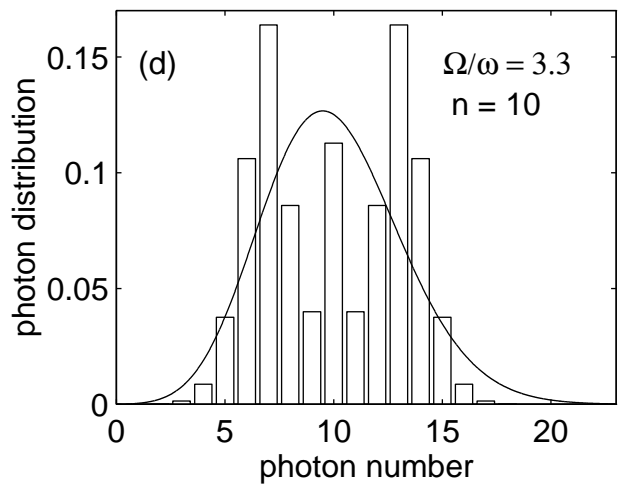
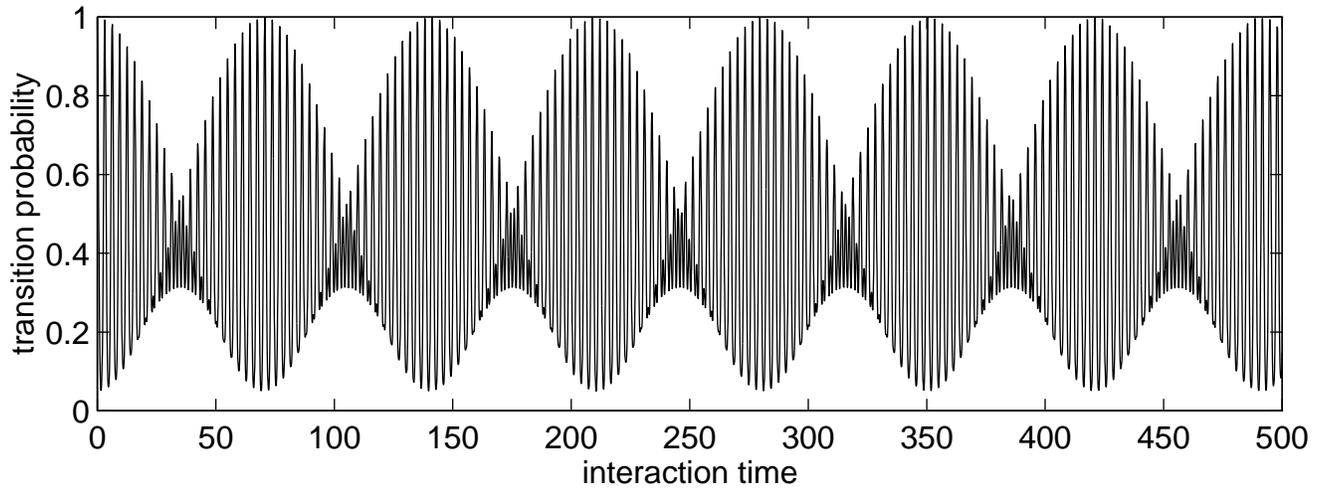
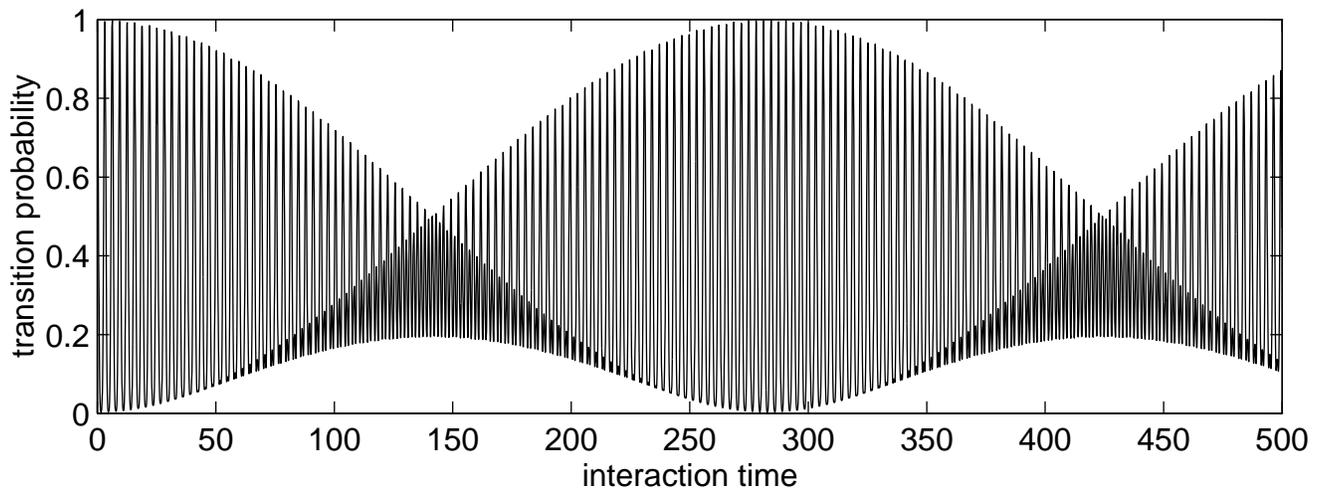
\end{document}